\documentclass[doublespacing]{elsart}

\usepackage{epsfig}

\usepackage{amssymb}

\begin{document}

\begin{frontmatter}



\title{On Bayesian Treatment of Systematic Uncertainties in Confidence Interval Calculation}


\author[label1,label2]{Fredrik Tegenfeldt}\ead{Fredrik.Tegenfeldt@cern.ch} and \author[label2]{Jan Conrad \corauthref{cor1}}\ead{Jan.Conrad@cern.ch}
\address[label1]{Iowa State University, Ames, IA 5011-3160, USA}
\address[label2]{CERN, PH-EP Department, CH-1211 Geneva 23}
\corauth[cor1]{Corresponding author: Jan Conrad, PH-EP Dept, F01910, CH-1211 Geneva 23}
\begin{abstract}
In high energy physics, a widely used method to treat systematic uncertainties in confidence interval calculations is based on combining a frequentist construction of confidence belts with a Bayesian treatment of systematic uncertainties. In this note we present a study of the coverage of this method for the standard Likelihood Ratio (aka Feldman \& Cousins) construction for a Poisson process with known background and Gaussian or log-Normal distributed uncertainties in the background or signal efficiency. For uncertainties in the signal efficiency of upto 40 \% we find over-coverage on the level of 2 to 4 \% depending on the size of uncertainties and the region in signal space. Uncertainties in the background generally have smaller effect on the coverage. A considerable smoothing of the coverage curves is observed. A software package is presented which allows fast calculation of the confidence intervals for a variety of assumptions on shape and size of systematic uncertainties for different nuisance parameters. The calculation speed allows experimenters to test the coverage for their specific conditions.

\end{abstract}


\end{frontmatter}


\section{Introduction}
The calculation of confidence intervals in case of presence of systematic uncertainties is an open problem. Systematic uncertainties are those uncertainties present in parameters which will affect the calculated confidence interval, but which are not of prime interest, so called {\it nuisance parameters.} Examples for nuisance parameters are the efficiency or the predicted background rate.\\
In 1992 Cousins \& Highland \cite{Cousins:1992a} proposed a method which is based on a Bayesian treatment of the nuisance parameters. They proposed to perform a frequentist construction (following Neyman \cite{Neyman:1937a}) and to replace the probability density function (PDF) describing the statistical process of prime interest with a PDF which is obtained by a convolution the original PDF with the one describing the uncertainties in the nuisance parameters:
\begin{equation}
P(n|s) \longrightarrow \int P(n|s,\epsilon') P(\epsilon'|\epsilon)d\,\epsilon'
\end{equation}
where $\epsilon'$ is the true value of the nuisance parameter, $\epsilon$ denotes its estimate and $s$ and $n$ symbolize the signal hypothesis and the experimental outcome respectively.\\
Highland \& Cousins only treated the case of Gaussian uncertainties in the signal efficency. The method has since been generalized to operate with the modern unified ordering scheme proposed by Feldman \& Cousins \cite{Feldman:1998a} and taking into account several nuisance parameters (efficiencies and background) and correlations \cite{Conrad:2002kn}. This generalized method has already been  used in a number of particle and astroparticle physics experiments (e.g. \cite{LIGO} \cite{BELLE} \cite{KAMLAND}, \cite{HERAB}, \cite{AMANDA}).\\ 
The most crucial property of methods for confidence interval construction is the coverage, which states: \\
\\
{\it A method is said to have coverage (1-$\alpha$) if, in infinitely many repeated experiments the resulting confidence interval includes the true value with probability (1-$\alpha$) irrespective of what the true value is.}\\
\\
100(1-$\alpha$)\% is hereby commonly taken to be 68\%  90\%, 95\% etc. Recently, the coverage properties of a fully frequentist method, the Profile Likelihood method, have been studied \cite{Rolke:2004mj}. The Profile Likelihood method was found to have surprisingly good coverage properties with a small (mostly negligible) amount of undercoverage.\\
In this note we undertake a systematic study of the coverage of the Bayesian method in order to give more substance to further recommendations on what method to use to calculate confidence intervals in presence of systematic uncertainties. Previous studies \cite{Conrad:2002ur} of this method dealt only with certain limiting cases and were constrained by computational requirements.\\
The note is organized as follows: in the next section we will review the frequentist construction of confidence intervals, in particular the likelihood ratio ordering scheme. We will describe the Bayesian method to incorporate systematics in section \ref{sec::sys}. The C++ library used to perform the calculations is described in section \ref{sec::pole}. Coverage tests and connected subtleties are discussed in section \ref{sec::cov}. The final section is devoted to discussion and conclusion.

\section{Frequentist confidence interval construction}
Let us consider a Poisson probability density function (PDF), $p(n)_{s+b}$, for a fixed but unknown signal, $s$, in the presence of
a known background with mean $b$. For every value of
$s$ we can find two values $n_1$ and $n_2$ such that 
\begin{equation}
\sum_{n'=n_1}^{n_2} p(n')_{s+b} = 1 - \alpha
\label{eq:neyman0}
\end{equation}
where $1-\alpha$ denotes the confidence level (usually quoted as 
a 100(1-$\alpha$)\% confidence interval).
Since we assume a Poisson distribution, the equality will generally
not be fulfilled exactly.  A set of intervals
$[n_1(s+b,\alpha),n_2(s+b,\alpha)]$ is called a {\em confidence
belt\/}. Graphically, upon a measurement, $n_o$, the {\em confidence interval\/} $[s_1,s_2]$ is determined by the intersection of the vertical line
drawn from the measured value $n_o$ and the boundary of the confidence
belt. The probability
that the confidence interval will contain the true value $s$ is $1 -
\alpha$, since this is true for all $s$ per construction. 
The choice of the $n_1$ and $n_2$ is, however, not unique to define the
confidence belt.  An additional criterion has to be applied. 
The currently recommended ordering scheme ~\cite{Feldman:1998a} ~\cite{Kendall:91} of the elements in the sum in equation (\ref{eq:neyman0}) is based on likelihood ratios. This approach automatically provides central confidence intervals when motivated and upper limits when necessary, therefore it is often denoted as the ``unified approach''. The following algorithm is applied in solving equation (\ref{eq:neyman0}):\\
For each $n$ the $s_{best}$ is found which maximizes the
likelihood $\mathcal{L}(n)_{s+b}$. In case of a simple Poisson
distribution with known background, $s_{best}$ is given by $max(0,n-b)$.
Then for a fixed $s$ the ratio
\begin{equation}
R(s,n)_{\mathcal{L}} = \frac{\mathcal{L}_{s+b}(n)}{\mathcal{L}_{s_{best}+b}(n)}
\end{equation}
is computed for each $n$, and all $n$'s are consequently ranked
according to the value of this ratio. Values of $n$ are included in
the confidence belt starting with the $n$ with the 
highest rank (largest $R_{\mathcal{L}}$) and then decreasing rank 
until $\sum_{n=n_1}^{n_2} p(n)_{s+b} = 1 - \alpha$. 
After the confidence belt has been constructed in this way, the 
confidence interval $[s_1,s_2]$ is found as described above.  
\section{Incorporation of systematic uncertainties: the Bayesian way}
\label{sec::sys}
The ordering schemes are unaffected by the way of treating systematic uncertainties considered here. As mentioned earlier, the PDF describing the statistical process will however be modified. Two concrete examples are the following:  In the case that the only uncertainty present is a theoretical (assumed Gauss-shaped) uncertainty of the background process the PDF is modified to:
\begin{equation}
q(n)_{s+b} = \frac{1}{\sqrt{2\pi}\sigma_{b}}\intop_0^{\infty}p(n)_{s+b'}\;\;e^{-\frac{(b-b')^2}{2\sigma_b^2}}\,db'
\label{eq:b}
\end{equation}
 Here $b$ is the estimated background level, 
and $\sigma_b$ is the uncertainty in the background estimation. 
If, in addition to the theoretical uncertainty for background, there
is the need to include the uncertainty in the signal detection efficiency the expression for $q(n)_{s+b}$ might be
extended to:
\begin{equation}
\begin{array}{ll}
q(n)_{s+b} =  
\frac{1}{2\pi\sigma_{b}\sigma_{\epsilon}} \times & \\
\intop_0^{\infty}\intop_0^{\infty}p(n)_{b'+ 
\epsilon's}\;\;e^{\frac{-(b-b')^2}{2\sigma_b^2}}\;\;
e^{\frac{- (1 -\epsilon')^2}{2\sigma_{\epsilon}^2}}db'd\epsilon' &\\
\end{array}
\label{eq:epsilon}
\end{equation}
where $\sigma_{\epsilon}$ is the uncertainty in the detection
efficiency expressed in {\em relative\/} terms with respect to 
the nominal efficiency\footnote{of course the efficiency could be defined in absolute terms as well.}. 
It is important to realize that the integration variables, here $\epsilon'$ and $b'$, are the possible ``true'' (but unknown) values of nuisance parameter. This indicates that this method is based on Bayesian statistics.\\
Some examples for confidence intervals computed by this method are shown in table \ref{tab::sys}.

\section{pole++}
\label{sec::pole}
For the coverage studies presented in this paper a reasonably fast and efficient code is required. Hence, a user-friendly and flexible C++ library of classes was developed based on the FORTRAN routine presented in \cite{Conrad:Pole}. The library is independent of external libraries and consists of two main classes, {\em Pole} and {\em Coverage}. The first class takes as input the number of observed events, the efficiency and background with uncertainties and calculates the limits using the method described in this paper. The integrals are solved analytically. {\em Coverage} generates user-defined pseudoexperiments and calculates the coverage using {\em Pole}. Presently the library supports Gauss and log-Normal distributed PDF for description of the nuisance parameters. Flat and used-defined PDFs are about to be implemented as well as correlations for the Gauss case. The class is dynamically optimized depending on if one wishes to calculate single (or few) confidence intervals or if one wants to perform a coverage study. Without these optimisations the calculation of a single interval takes about 1 second (wall clock time) on a 1 GHz Pentium III processor. The duration of a full coverage study (typically requiring the calculation of $\mathcal{O}(10^5)$ confidence intervals) ranges between a couple of minutes for small uncertainties and small signal hypotheses to order of 10  hours for large uncertainties and high signal hypotheses. The perfomance of a coverage study thus seems feasable for the particular set of systematic uncertainties that may appear in real experiments. The pole++ library can be obtained from http://cern.ch/tegen/statistics.html

\section{Coverage Studies}
\label{sec::cov}
\begin{figure}[c]
\epsfig{file=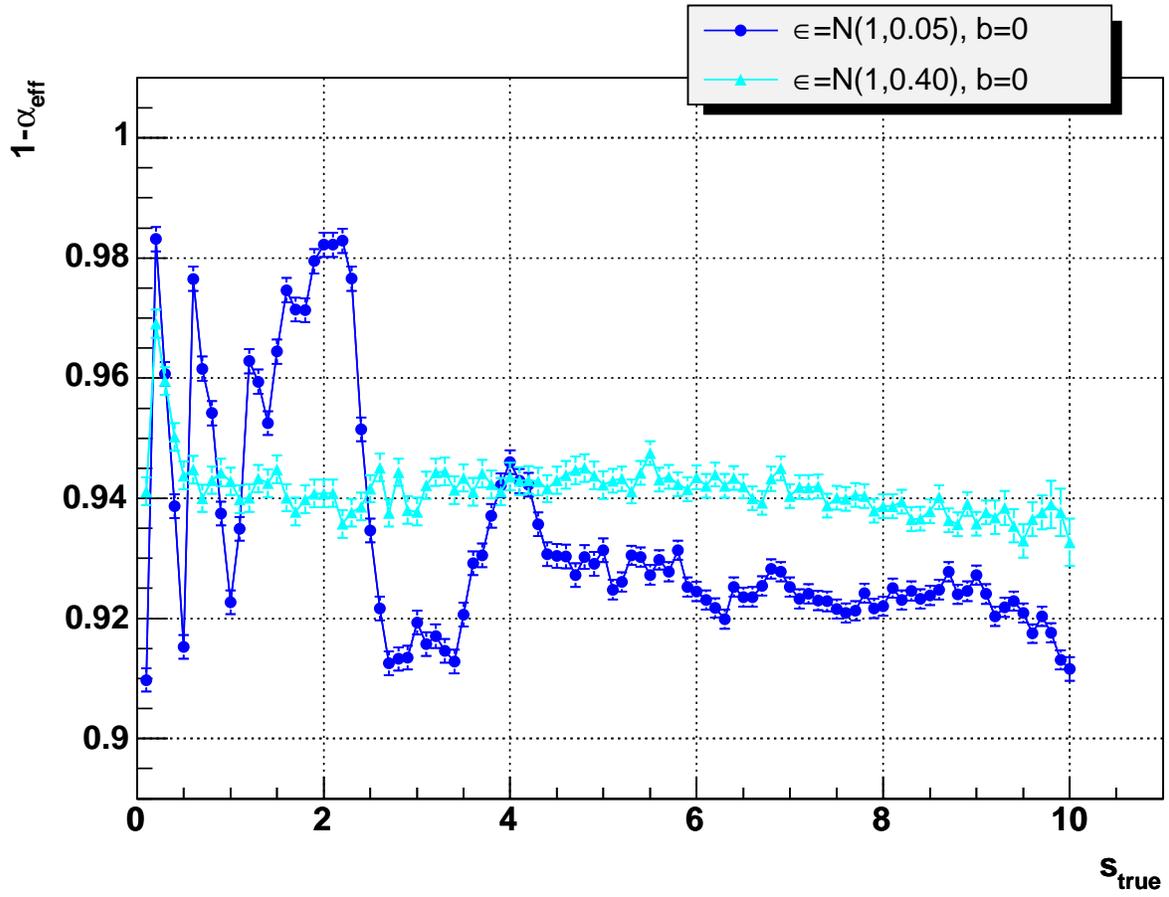,height=12cm} 
\caption{Calculated coverage as function of signal hypothesis. Two case are shown: 5 \% and 40 \% Gaussian uncertainties in the signal efficiency. The nominal coverage was 90\%.}%
\label{fig::sigeff}
\end{figure}

\begin{figure}[c]
\epsfig{file=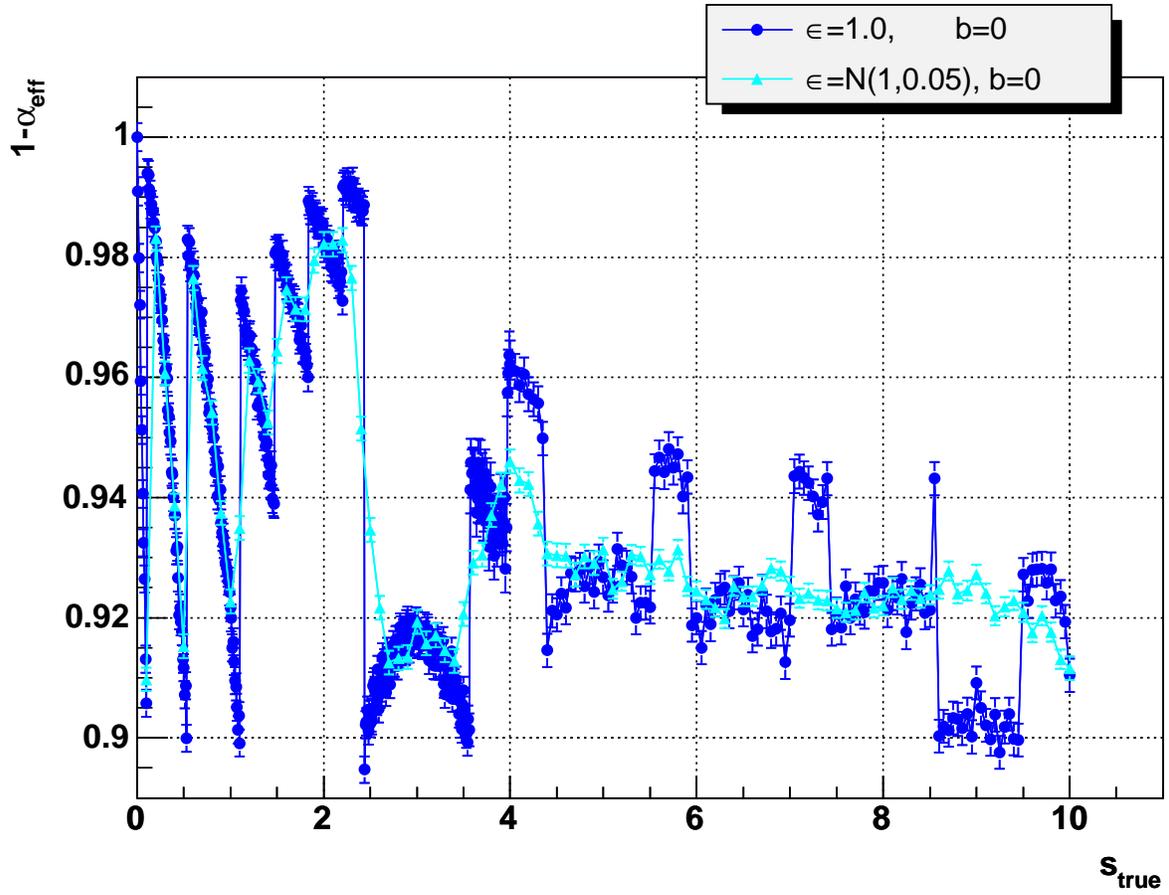,height=12cm} 
\caption{Calculated coverage with 5 \% systematic uncertainties is compared with the zero uncertainties case. The nominal coverage was 90 \%.}%
\label{fig::nounc}
\end{figure}

\begin{figure}[c]
\epsfig{file=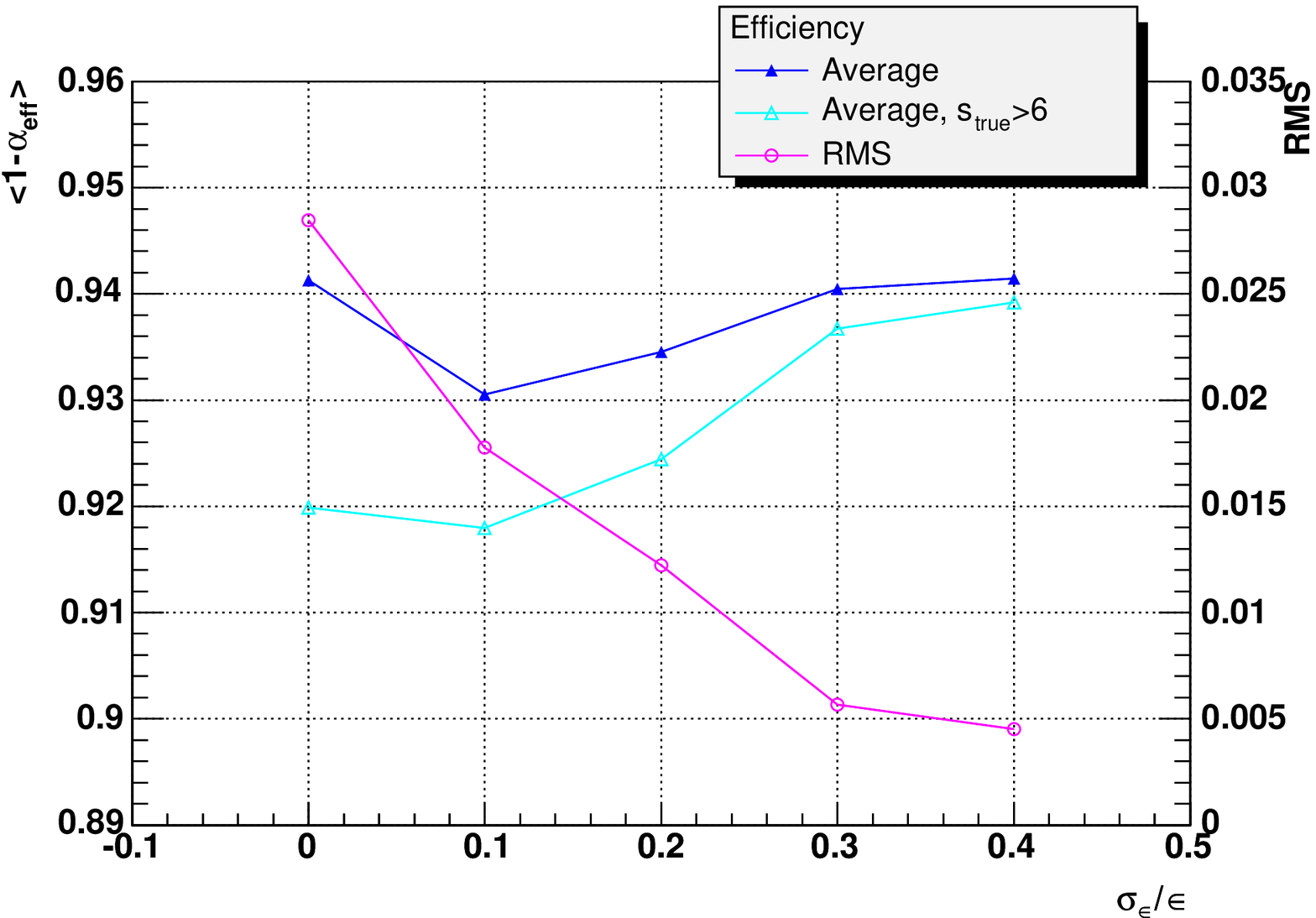,height=12cm}
\caption{Calculated mean coverage and rms as function of Gaussian shape uncertainties in the signal efficiency. The mean was calculated over the full range of signal hypotheses and for signal hypotheses larger than 6, respectively. The rms was calculated over the full range of signal hypotheses. The nominal coverage was 90 \%.}%
\label{fig::meanc}
\end{figure}

\begin{figure}[c]
\epsfig{file=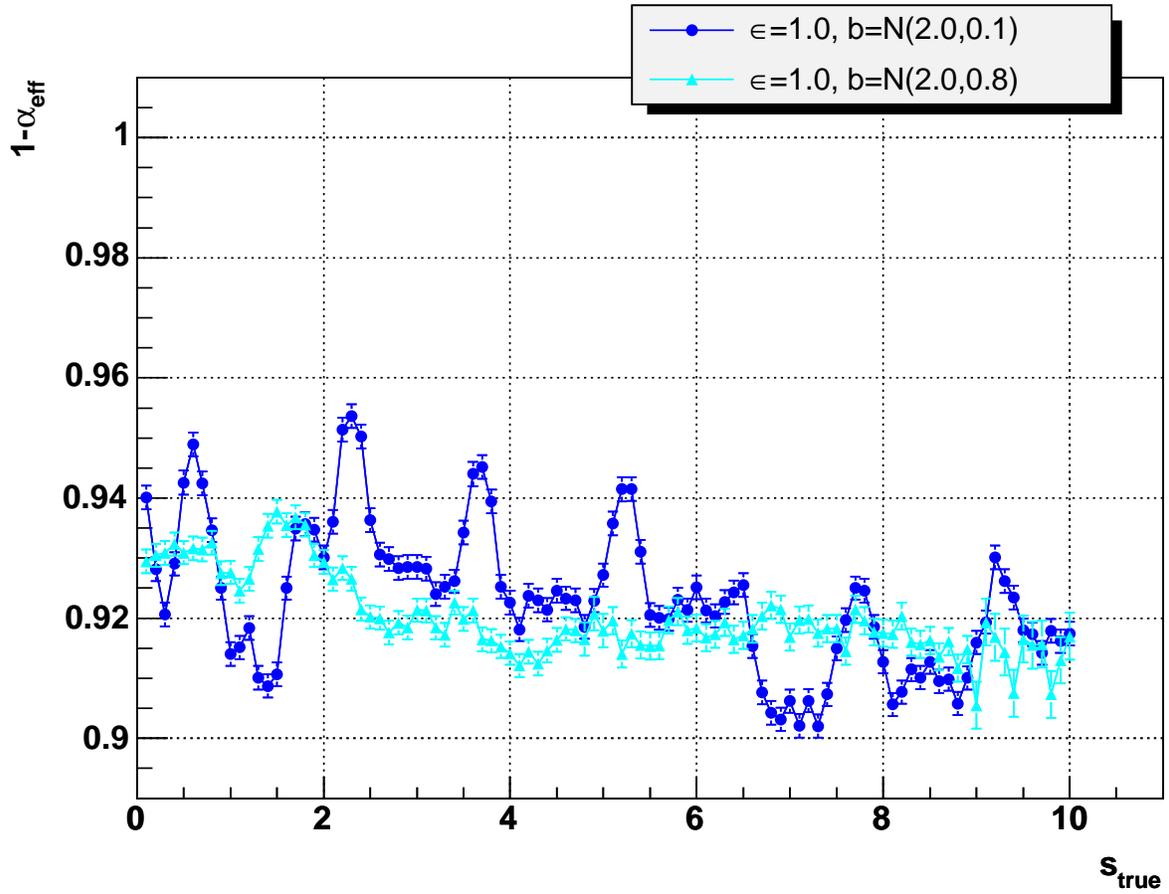,height=12cm} 
\caption{Calculated coverage as function of signal hypothesis. Two case are shown: 5 \% and 40 \% Gaussian uncertainties in the background prediction. The nominal coverage was 90 \%.}%
\label{fig::bgeff}
\end{figure}

\begin{figure}[c]
\epsfig{file=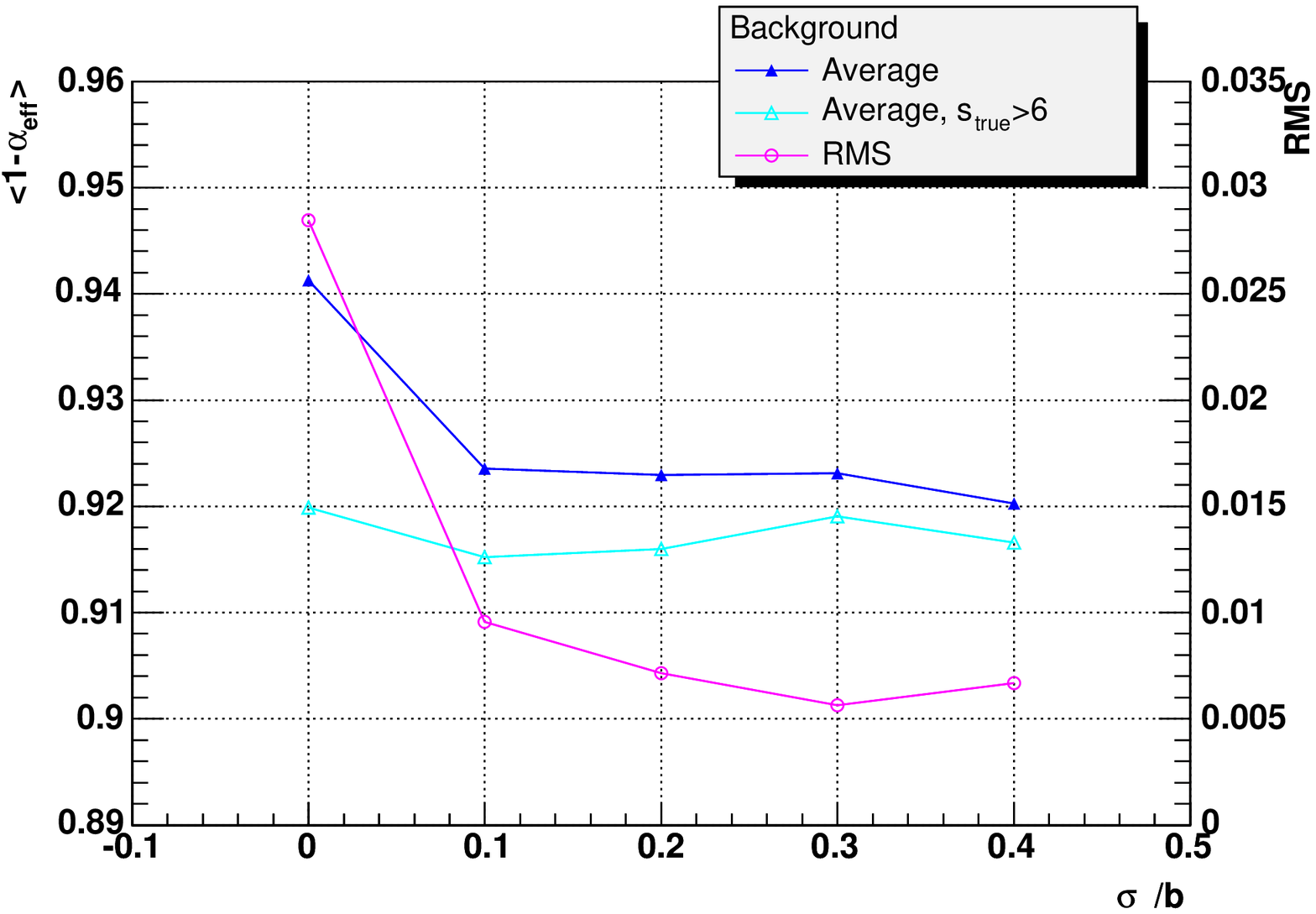,height=12cm}
\caption{Calculated mean coverage and rms as function of Gaussian shape uncertainties in the background prediction. The mean was calculated over the full range of signal hypotheses and for signal hypotheses larger than 6, respectively. The rms was calculated over the full range of signal hypotheses. The nominal coverage was 90 \%}%
\label{fig::meanb}
\end{figure}

\begin{figure}[c]
\epsfig{file=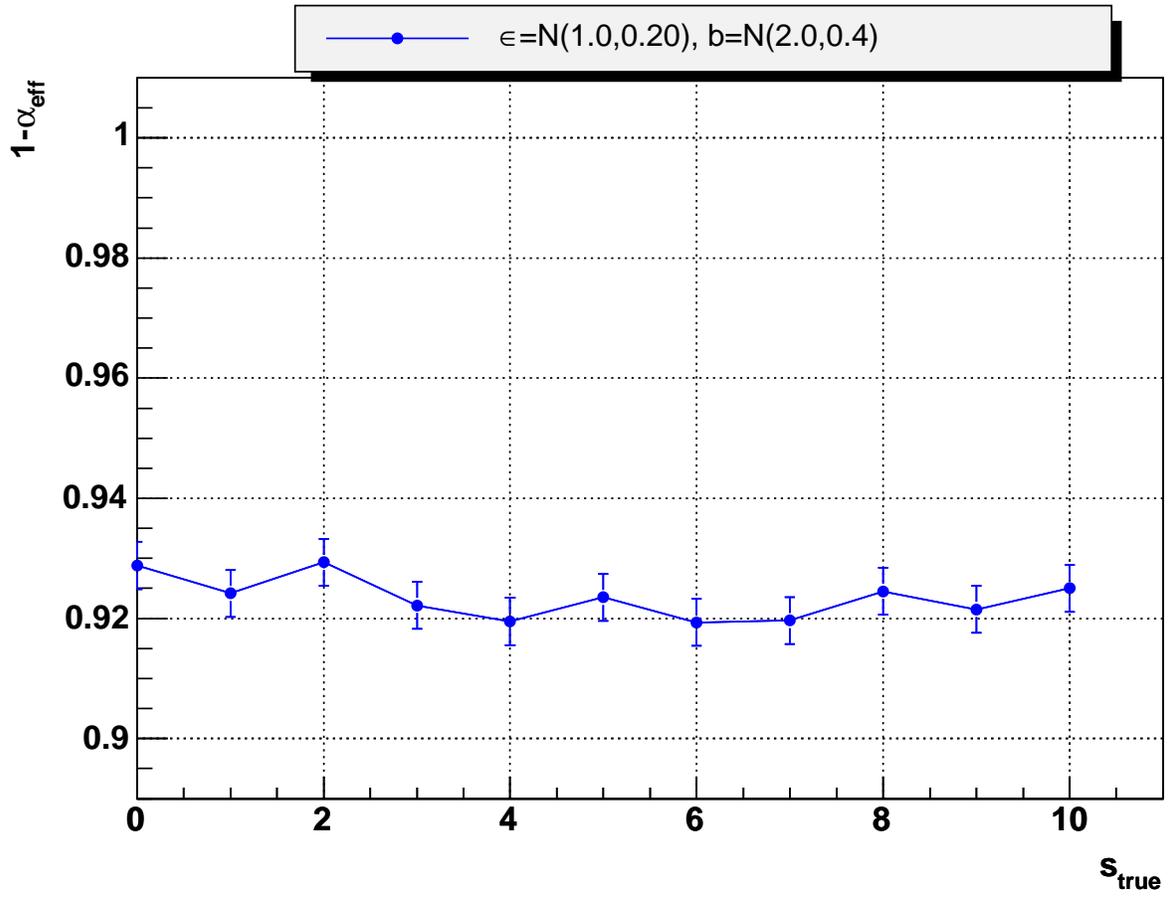,height=12cm} 
\caption{Coverage for the case that uncertainties are present in both the signal detection efficiency and in the background prediction. The nominal coverage was 90 \%}%
\label{fig::bgcor}
\end{figure}

\begin{figure}[c]
\epsfig{file=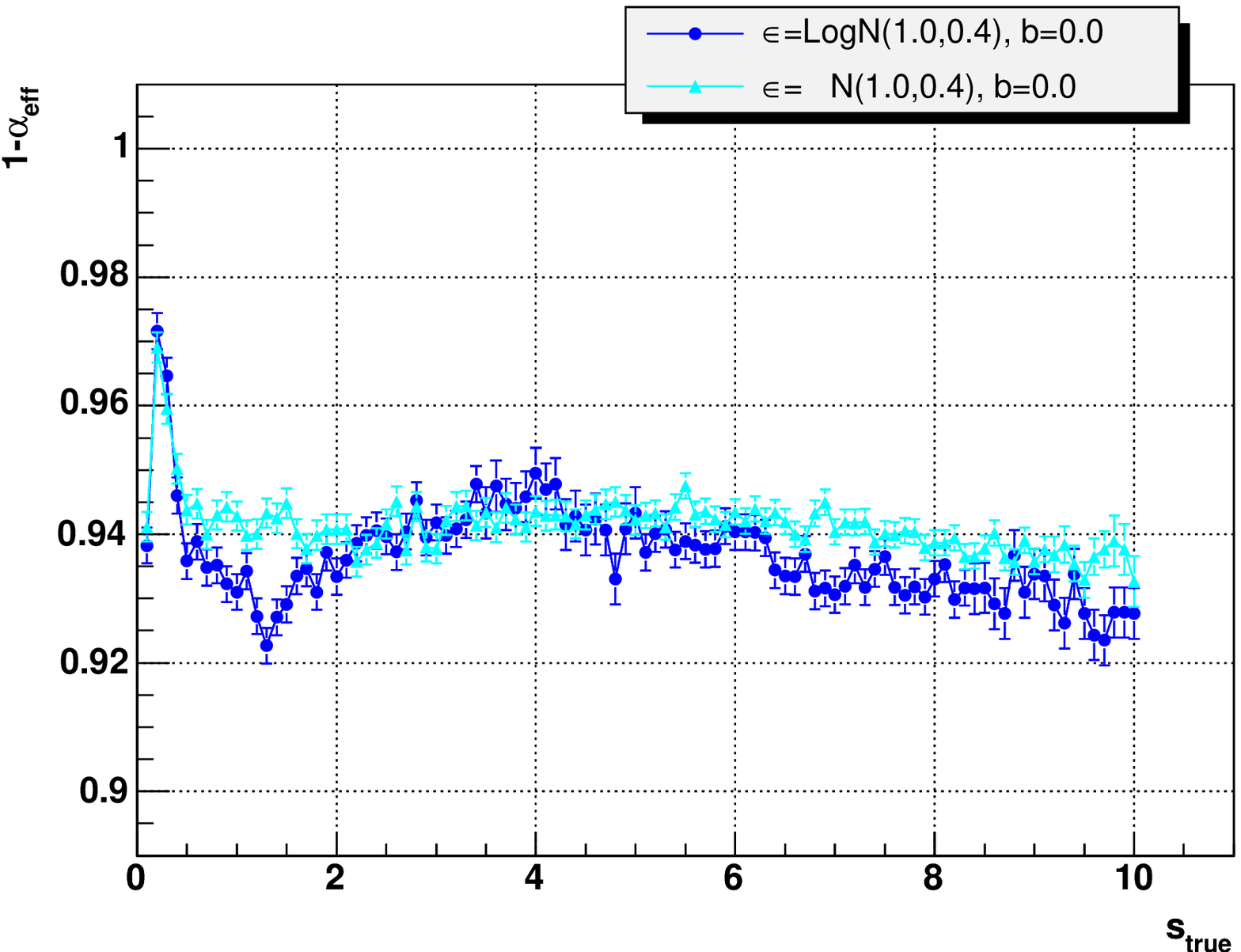,height=12cm} 
\caption{Coverage as a function of signal hypothesis for a Gaussian PDF and a log Normal PDF}%
\label{fig::logN}
\end{figure}

The coverage of the method is studied using MC simulations of pseudo-experiments with given true value of the prime parameter (signal) or nuisance parameter (efficiency, background). The {\it estimated} values of the nuisance parameters were assumed to be Gaussian or log-Normal distributed around the  given true value. The outcome of one experiment thus consisted of a number of observed events (following a Poisson distribution with known background and depending on the true efficiency and true background) and the estimate of the nuisance parameter(s).\\
Figure \ref{fig::sigeff} shows the coverage\footnote{we will denote the coverage calculated by the MC simulations as $(1-\alpha)_{eff}$ to distinguish it from the nominal coverage $(1-\alpha)$} as a function of signal hypothesis for two different sizes of uncertainties (5 \% and 40 \%). The uncertainty considered is in the signal efficiency and assumed to be Gauss-shaped.  It can be seen that the Bayesian method causes over-coverage which is larger for larger systematic uncertainty. This is also reflected in figure \ref{fig::meanc} where the mean coverage (mean taken over all tested signal hypotheses) is shown as a function of assumed uncertainty.  For Gaussian uncertainties in the signal efficiency we find an increase in mean coverage by $\sim$ 1\% in the uncertainty range between 10 \% and  40 \%.\\
The see-saw structure which is generally seen in the no-uncertainty case (due to the discrete experimental outcome) is considerably smoothened when systematic uncertainties are introduced. This is exemplified in figure \ref{fig::nounc}, which  compares the coverage curve with 5 \% uncertainties on the signal efficiency with the zero uncertainties case. To quantify this further we present the rms of the coverage together with the mean in figure \ref{fig::meanc}. The smoothing is due to the fact that we add a continuous variable to the problem, meaning there are more degrees of freedom to fulfil the sum condition of equation \ref{eq:neyman0}.
The effect of the smoothing is that whereas for some signal hypothesis the coverage is increased for others it is decreased with respect to the zero uncertainty case. The mean coverage is therefore only rather weakly dependent on the systematic uncertainties. For higher signal hypotheses, where the effect of the smoothing is less pronounced, the effect of the uncertainties is therefore stronger. Taking the mean over signal hypotheses $s_{true}>$6 the mean coverage increases from 92 \% (at zero uncertainties) to 94 \% (at 40 \% uncertainties). A side effect is that the introduction of rather small uncertainties seems to improve the coverage with respect to the zero uncertainty case for parts of the tested hypothesis space (see e.g. figure \ref{fig::nounc}).\\
Figure \ref{fig::bgeff} shows the coverage for two different sizes of Gaussian uncertainty on the background estimate. Except for the smoothing effect the coverage is seemingly independent of the size of the uncertainties. An uncertainty in the background will yield similar results to an uncertainty in the signal efficiency only in the regime were the signal hypotheses are of similar size as the background expectation. For larger signal hypotheses the coverage curve for the uncertainty in the background will be approaching the zero uncertainty case, which is reflected in a slight slope of the coverage curve for high uncertainty in figure \ref{fig::bgeff}. The result is thus that a mean coverage depending on uncertainties in the background will be less affected than the corresponding curve with uncertainties in the signal efficiency. We show the mean coverage as well as the rms in figure \ref{fig::meanb}.\\
In figure \ref{fig::bgcor} the effect of having to consider uncertainties both in signal and background detection efficiency is visible.  At large signal hypotheses the coverage plot is dominated by the uncertainty in signal efficiency, at low signal hypotheses coverage benefits from the smoothing effect, since we added yet another degree of freedom.\\
It should be noted that for large uncertainties the Gaussian model is not appropriate. 
 The reason is that a Gaussian model will then significantly extent to the unphysical region of the space of experimental outcomes. In the Bayesian treatment this case is dealt with by truncating the Gauss distribution at zero \footnote{it is worth noting that Cousins \& Highland develope their method for unbiased estimators, i.e. the problem would not be treated in this way.}, and this is consequently the way it is dealt with during the coverage test. 
Considering the coverage test there is an additional subtlety: in order for the measured efficiency to be a maximum likelihood estimate of the true efficiency we would have to renormalize the Gauss distribution to the truncated Gauss distribution in order to obtain the correct PDF. 
However, instead of doing this a posteriori fix, it is more reasonable to use a log-Normal distribution to model the uncertainty in the efficiency. The coverage for the log-Normal  distribution is shown in figure \ref{fig::logN}. As can be seen for the highest uncertainties considered in this note, the Gauss distribution is still a very good approximation to the log Normal model.
\section{Discussion and Conclusions}
\label{sec::dis}
In this note we presented coverage studies for the Bayesian treatment of systematic uncertainties.  
One overall conclusion is that the Bayesian treatment leads some over-coverage. However, introducing a continuous nuisance parameter into the discrete Poisson problem results in a smoothing of the coverage curves. The mean coverage is therefore only weakly affected by the Bayesian treatment of nuisance parameters and under certain circumstances even improves with respect to the zero uncertainty case.\\
In a frequentist approach the meaured estimate of a nuisance parameter is considered to be distributed around a given true value, which is consequently the way coverage was calculated. The Bayesian method on the other hand views the true value as distributed around the measured value. The underlying assumption for going from one approach to the other is (at least in case of a Gauss-distribution) a flat prior probability of hypotheses. The present study indicates that this assumption does not lead to a violation (except for over-coverage) of the coverage requirement.\\
The routines used for the presented calculations are reasonably fast and publicly available. A coverage study is therefore feasable for each problem at hand. The confidence level required for the confidence intervals could then in principle be adjusted to recover correct coverage.
\section{Acknowledgements}
The authors thank Robert Cousins for a very useful discussion.

\section{Figures and tables}
\begin{table}[htbp]
\begin{center}
\begin{tabular}{|l|l|l|l|l|}
\hline\hline
$n_0$ 	&$b$ &rel. signal efficiency & Likelihood Ratio \vspace{-0.1cm}\\
	&    &uncertainty      &interval     \\ \hline
2     	&2       &0               &0: 3.91     	\\
     	&        &0.2             &0: 3.89      \\
      	&        &0.4             &0: 4.68	\\ \hline
4     	&2       &0               &0: 6.59      \\
      	&        &0.2             &0: 7.16	\\
      	&        &0.4             &0: 8.99	\\ \hline
6       & 2      &0               &1.08: 9.47   \\
        &        &0.2             &1.07:10.09   \\
        &        &0.4             &1.02:13.31    \\  \hline\hline   
\end{tabular}
\caption{\label{tab::sys}Examples of likelihood ratio 90\% confidence intervals with Bayesian treatment of systematic uncertainties. Uncertainties are assumed to be Gauss distributed in the signal efficiency.}
\end{center}
\end{table}

\begin{thebibliography}{00}
\bibitem{Cousins:1992a} R.~D.~Cousins and V.~ L.~Highland, Nucl. Instrum. Meth. {\bfseries A320}, 331, (1992).
\bibitem{Neyman:1937a} J.~Neyman, Phil. Trans. Royal Soc. London {\bfseries A}, 333, (1937).
\bibitem{Feldman:1998a} G.~J.~Feldman and R.~D.~Cousins, Phys. Rev {\bfseries D57},  3873, (1998).
\bibitem{Kendall:91} A.~Stuart and J.~K.~Ord: Kendall's Advanced
Theory of Statistics, Vol. 2, Classical Inference and Relationship,
Oxford University Press, New York (1991).
\bibitem{Conrad:2002kn}
J.~Conrad, O.~Botner, A.~Hallgren and C.~Perez de los Heros,
Phys.\ Rev.\ D {\bf 67} (2003) 012002
[arXiv:hep-ex/0202013].
\bibitem{LIGO}
B.~Abbott {\it et al.}  [LIGO Collaboration],
Phys.\ Rev.\ D {\bf 69} (2004) 102001.
\bibitem{BELLE}
Y.~Chao {\it et al.}  [Belle Collaboration],
Phys.\ Rev.\ D {\bf 69} (2004) 111102
[arXiv:hep-ex/0311061].

\bibitem{KAMLAND}
K.~Eguchi {\it et al.}  [KamLAND Collaboration],
Phys.\ Rev.\ Lett.\  {\bf 92} (2004) 071301
[arXiv:hep-ex/0310047].

\bibitem{HERAB}
I.~Abt {\it et al.}  [HERA-B Collaboration],
arXiv:hep-ex/0405059.
\bibitem{AMANDA}
J.~Ahrens  [AMANDA Collaboration],
Phys.\ Rev.\ Lett.\  {\bf 92} (2004) 071102
[arXiv:astro-ph/0309585].
\bibitem{Rolke:2004mj}
W.~A.~Rolke, A.~M.~Lopez, J.~Conrad and F. James
arXiv:physics/0403059.
098301, (2000).
\bibitem{Conrad:2002ur} J.~Conrad, O.~Botner, A.~Hallgren and C.~Perez de los Heros, published in Proc. of Conference on Advanced Statitical Techniques in Particle Physics, Durham, England, March 2002
\bibitem{Conrad:Pole}  J.~Conrad, Computer Physics Communications {\bfseries 158} 117-123 (2004)

\end{thebibliography}
\end{document}